\documentclass[docs]{article}
\begin{document}
\title{The Physics of Flavor -- Challenge for the Future}
\author{Harald Fritzsch\\
\textit{\normalsize{Ludwig-Maximilians-Universit\"at M\"unchen, Sektion Physik,}}\\
\vspace{-0.6cm}\\
\textit{\normalsize{Theresienstra\ss e 37, D-80333 M\"unchen, Germany}}}
\maketitle
\begin{abstract}
This is the summary talk for the theoretical part of ICFP03. The contents
of the talk are reviewed, and a general outlook is given.
\end{abstract}
It is a special pleasure for me to give the Summary Talk on Theory at the
conference on flavor physics, held in Seoul (Korea) in 2003. Not far from here,
in China, one talks about philosophers like Konfuzius or Laotse. The physics
of flavor is certainly not the world of Konfuzius, based on rigid laws and
simple structures, but resembles the world of Laotse, full of complexity, a
world which is in the eternal swing between Yin and Yang.

In particle physics the physics of flavor is the field with the highest
complexity and the richest phenomenology, with phenomena ranging from the
spectra of mesons and baryons, from CP--violation, from weak decays up to
exotics phenomena like neutrino oszillation. It deserves its name from a
discussion I had with M. Gell--Mann in 1970, when I was a graduate student,
working with Gell--Mann at CALTECH. One day we stopped on the way back from
lunch in the Baskin--Robins place in Pasadena. When we had icecream, I
remarked that the 32 flavors, offered by Baskin and Robins, are much more
than the three flavors {\it u, d} and {\it s}, we are dealing with in case of
the quarks. Gell--Mann got hooked on this remark, referring to the physics of
flavor in the future, when we were dealing with the weak interaction of the
quarks, and not with the colors, reserved to the description of the strong
interactions.

Flavor physics emerged as an independent field of research in particle physics
only after the Standard Model had come up in its first contures shortly after
the beginning of the 70'ies. Today the physics of flavor is at the same time
the physics of a multitude of free parameters, or, better, of parameters,
which in the Standard Model can be regarded as free. If we leave out the
neutrino masses, we have 18 parameters, entering the Standard Model. Including
the masses of neutrinos and related phenomena, we have to add another 9
parameters, 3 for the neutrino masses, plus 6 parameters describing the mixing,
now altogether we are dealing with 27 parameters. Of particular interest for
flavor physics are the 12 masses of the fermions and the 10 parameters,
describing the mixing for quarks and leptons, four for the quarks and six for
the neutrinos, including three phases for the neutrinos, under the assumption,
that neutrinos are Majorana particles, i.e. not massive Dirac states.

Let me, at this point, make a remark about the Higgs particle, which remains
unseen thus far in the experiments. If we take in particular the LEP results,
we have to conclude that the mass of the Higgs particle is about 91 + 57 /
- 37 GeV. A large part of this mass range is, however, already excluded from
the direct search using the LEP machine: masses less than about 115 GeV are not
allowed. Nevertheless the $H$--particle should have a mass of less than
211 GeV (with 95 \% c.l.h), otherwise something is wrong with the Standard
Model. In the MSSM--model one of the {\it H}--particle should have a mass
less than 135 GeV.

Assuming that the Standard Model or its MSSM--extension are not totally wrong,
we would expect a rich harvest concerning ``Higgs physics'' with LHC, but
even more with a L.C. `(`Linear Collider''), reaching an energy of about 800
GeV, like the TESLA machine, under study at DESY in Germany.

Let me also make a few remarks about the fermion masses. The most surprising
feature about them is the very impressive mass hierarchy. The spectrum starts
presumably just below 1 eV (neutrino masses), and then it climbs up to the
{\it t}--mass of about 174 GeV, i. e. 174.000.000.000 e.V. What I also find
remarkable, is the fact that the quark masses are such that the ratios are
equal:
\begin{eqnarray}
m_c : m_t & \cong & m_u : m_c \nonumber \\
m_s : m_b & \cong & m_d : m_s
\end{eqnarray}

Many years ago, in 1987, I once gave a colloquium at CALTECH and talked about
these ratios. I also remarked that the {\it t}--quark would have to be very
massive if these ratios hold for the {\it u}--type quarks, with a mass around
170 GeV. For the talk I had placed the quark masses as straight lines on a
logarithmic plot. R. Feynman, who was at the colloquium, did not believe my
argument and remarked: ``You know, on a logarithmic plot even Sophia Loren will
look like a straight line.''

Everybody laughed, but I remarked, that experimentalists should find the
{\it t}--quark eventually, even with the high mass I was expecting, and they
found it, with a mass very close to the expected value.

Let me add that I find the way the fermion masses are introduced in the
Standard Model the least attraction feature of the model, furthermore a
feature, which has essentially no predictive power. The Higgs particles had
to be introduced to give masses to the $W$-- and $Z$--bosons. Whether it really
gives masses to the fermions, remains unclear. I guess
that nature found another way to introduce the fermion masses, and eventually
we will find out which way. But if the Higgs particle does not couple to the
fermions as expected in the Standard Model, it will also not decay to the
fermions with an amplitude proportional to the fermion mass. Thus alternative
decay modes might become relevant, like the
gluonic decay $H \rightarrow gg$ or the photonic decay
$H \rightarrow \gamma \gamma $. In Munich we have convinced our experimentalists
at LEP to look for such decays, but nothing was found\cite{CAL}.

In the Standard Model the mass of the {\it t}--quark is roughly comparable to
the mass scale, given by the Fermi constant, i. e. the v.e.v. of the
{\it H}--field: $v \cong 246$ GeV. However the observed {\it t}--mass is
close to $v / \sqrt{2} \cong 174$ GeV, i. e. we have
$v \cong \sqrt{2} \, m_t$. Personally I find this relation quite remarkable. It
might be an accident, but to think that the ratio $\sqrt{2}$ might have
something to do with a Clebsch--Gordan coefficient, related to an underlying
symmetry. Nobody has found such a symmetry thus far, but we should keep this
ratio $\sqrt{2}$ in mind.

I think the simple features exhibited by the fermion mass spectrum ask for a
deeper understanding, which goes much deeper than the shallow interpretation
given by the Standard Model. In fact, the mass spectrum of the fermions seems
to be a clear sign that there is a physics activity beyond the frontier line
drawn by the Standard Model, and presumably this line is not much below the
surface explored thus far in the experiments. The first experiments at the
LHC might open it up completely.

Let me come to the flavor physics, as discussed in this meeting. I like to do
this by first looking at the flag of the country. In the center of the flag are
two drops in red and blue, holding each other. They represent the two eternal
forces Yin and Yang. They are in permanent motion, but keep up harmony. I like
to compare this with the two sides of one field, represented by theory and
experiment. Indeed, flavor physics is the field in particle physics where
theory and experiment hold each other in balance, where theoretical progress
can soon be tested by experiment, and where new experimental facts are soon
understood by theory.

The Yin--Yang symbol in the center is surrounded by four elements, which like
the center represent unity. Above, on
the left, is KIEN, representing the creative element. I identify this with
the general flavor problem and the physics beyond the Standard Model.

Above, on the right, is KAN, the dangerous element. I identify this with the
field of weak decays and QCD, which can also be full of dangers. On the lower
side left is LI, the adhering element of life. I like to identify this with
neutrino physics, the new field of flavor physics, which has given us a lot of
new insights in the last two years.

The last element, on the lower right side, is KUN, the receiving element of
life. I like to identiy this with the field of flavor mixing and the masses
of the fermions, keeping in mind that this field might actually be the one
which opens soon the door to the physics beyond the Standard Model.

Before I make statements to the various theoretical talks at this conference
let me make a short remark about the way we describe the flavor mixing for
the quarks. Often this is done using  the Wolfenstein representation of the
CKM--matrix. In this representation the element $V_{us}$ is given by the
parameter $\lambda $, and the element $V_{cb}$ by $A \lambda^2$. I do not like
this representation, since I think that $V_{cb}$ and $V_{us}$ are different
elements, and I can easily think of a situation in which $V_{cb}$ is there,
but $V_{us}$ is zero -- this would not be possible in the Wolfenstein
description.

Xing and I studied some time\cite{Frit} ago the various ways to describe the flavor
mixing. We came to the conclusion that the ``standard'' parametrization as well
as the original Kobayashi--Maskawa representation\cite{Kob} were introduced without
taking possible links between the quark masses and the flavor
mixing parameters into account. Xing and I studied in particular a
parametrization, which I had introduced already in 1979\cite{Frit1}. It is a
description of the flavor mixing as an evolutionary process. In the limit in
which all masses, except $t$ and $b$, are zero, there is no flavor mixing. Once
the $c-s$ masses are introduced, the flavor mixing is reduced to a mixture
between the $(c, s)$--system and the $(t, b)$--system, described by an angle
$\Theta $.

As soon as the $(u, d)$--masses are introduced, the full flavor mixing matrix
involving a complex phase parameter and two mixing angles $\Theta_u$ and
$\Theta_d$ appears. These two angles can be interpreted as rotations between
$(u, c)$ and between $(d, s)$. The mixing matrix is given by
\begin{equation}
V = \left( \begin{array}{ccc}
c_u & s_u & 0 \\
-s_u & c_u & 0 \\
0 & 0 & 1
\end{array} \right) \left( \begin{array}{ccc}
e^{-i \varphi}_u & 0 & 0 \\
0 & c & s \\
0 & -s & c
\end{array} \right)
\left( \begin{array}{ccc}
c_d & -s_d & 0 \\
s_d & c_d & 0 \\
0 & 0 & 1
\end{array} \right)  
\end{equation}
\\
where $s_u = sin \Theta_u, c_u = cos \Theta_u$ etc. The phase $\varphi $
describes CP--violation.

I like to note that in many models for the quark masses exist simple relations
between the mass eigenvalues and the angles $\Theta_u, \Theta_d$:
\begin{eqnarray}
tan \Theta_u & = & \mid V_{ub} / V_{cb} \mid \approx \sqrt{m_u / m_c}
\nonumber \\
tan \Theta_d & = & \mid V_{td} / V_{t s} \mid \approx \sqrt{m_d / m_s}
\end{eqnarray}
Furthermore simple ways to arrive at these relation predict also:
$\varphi = 90^{\circ}$.

Let me take, as an example, the mass values, normalized at 1 GeV:
$m_u = 6.8 MeV, m_d = 8.1 MeV, m_s = 190 MeV$ $m_c = 1050 MeV$.

In this case we have:
\begin{eqnarray} 
\mid V_{us} \mid^2 & \cong & \frac{m_d}{m_s} + \frac{m_u}{m_c} = 0.2202
\nonumber \\
tan \Theta_u & = & \sqrt{\frac{m_u}{m_c}} = \mid \frac{V_{ub}}{V_{cb}}
\mid \cong 0.081 \nonumber \\
tan \Theta_? & = & \sqrt{\frac{m_d}{m_s}} = \mid \frac{V_{td}}{V_{ts}} \mid
\cong 0.207
\end{eqnarray}

The triangle formed by $s_u, s_d$ and $V_{us}$ is congruent to the unitarity
tirangle. The angle $\alpha $ is given by $\varphi $, i. e. 90$^{\circ}$.
Furthermore we have $tan \beta = s_u / s_d \cong \sqrt{\frac{m_u}{m_c}} /
\sqrt{\frac{m_d}{m_s}} \approx 0.38,
sin 2 \beta \cong 0.68, \beta \approx
22^{\circ}$.

Thus the unitarity triangle would be given by:
\begin{equation}
\alpha \cong 90^{\circ}, \beta \cong 22^{\circ}, \gamma \cong 68^{\circ} \, .
\end{equation}
Thus far the experimented data are not in disagreement with these expectations.
The future will tell.

From Wu we heared that today the best value for $sin 2 \beta $ is:
\begin{equation}
sin 2 \beta = 0.736 \pm 0.049 \, .
\end{equation}
He emphasized that the use of isospin in studying charmless $B$--decays is ok,
but we have to be rather careful with the $SU(3)$--symmetry. Violations could
be large and are typically difficult to estimate. $SU(3)$--violations also do
affect the strong phases. Such violations may be responsible for apparent
inconsistencies between theory and experiment.

Du discussed two--body decays: $B \rightarrow PP, B \rightarrow PV$. In
particular he concentrated on the study of factorization. His estimate
$\gamma \approx 79^{\circ}$ looks a bit too large in comparison to the value
of $\gamma $, I have quoted above, but the error is large.

We heared about the problem related to the measurement of $B \rightarrow
\Phi + K_s$. Theoretically we would expect that this decay is quite similar to
the decay $B \rightarrow J / \psi + K_s$. For the disagreement between
experiment and the Standard Model (about 3.5 $\sigma $) I have no explanation.
We have to measure it again, also at Babar, and then have to see.

Sanda discussed two--body decays like $B^+ \rightarrow \eta' K^+, B^+
\rightarrow \pi^0 K^+$. I like to emphasize that $B$--decays involving the
$\eta'$--particle are of special interest, due to the large valence of the
$\eta '$ to gluons. Gluonic contribution may be responsible for making the
$\eta'$--decay particularly strong, and this may explain why the observed
$\eta '$--decay is so large\cite{Frit2}. In any case I see no need to involve new physics
beyond the Standard Model to explain this decay.

Won emphasized the importance to study the leptonic decays $B^+ \rightarrow
\mu^+ \nu, B^+\rightarrow \tau^+ \nu $ in searching for new physics. Certainly
any significant departure from the theoretical expectation would be a
signal for new physics.

Kang discussed the various ways to calculate the CKM mixing matrix in terms of
the fermion masses. He kept up the NNI--flag as a possibility, including
non--Hermitean matrices. He also emphasized the importance of the string theory
for understanding the flavor structure and the fermion masses. I still have
problems with this approach, since thus far string theory did not make any
contribution to the physics of flavor. In fact, string theorists hardly
understood anything in flavor physics. Of course, if one believes that string
theory eventually will explain anything, the physics of flavor might also come
out of string theory. But thus far I am sceptical about this possibility.

Yoshikawa discussed large electroweak penguin diagrams and their possible
relevance for new physics.As far as neutrinos are concerned, it was fascinating
to hear the various resports on neutrino oscillations, which are summerized
by Kleinknecht. In particular I like to emphasize the importance to look for
the transition $\nu_{\mu} \rightarrow \nu_e$ (i. e. the component $U_{e3}$).
We know that it is small, but the question remains how big or small it actually
is.

In general it is surprising to me that the neutrino mixing angles are large.
I like to mention that this feature was expected by a few theorists. In 1996
I worked on this with Z. Xing. One consideration was based on a simple
observation. If we start from a mass matrix for the charged leptons as
\begin{equation}
M = {\rm const.} \, \, \, \left( \begin{array}{ccc}
  		 		 	   	  1 & 1 & 1 \\
						  1 & 1 & 1 \\
						  1 & 1 & 1
						  \end{array} \right) \, 
\end{equation}
which has simple properties (the basic group is $S(3) \times S(3)$). One
obtains a strong hierarchical pattern for the charged lepton, i. e. ($0, 0,$
const.). At the same time the Majorana mass matrix could be
\begin{equation}
M = {\rm const.} \, \, \, \left( \begin{array}{ccc}
  		 		 	   	  		 1 & & \\
								 & 1 & \\
								 & & 1
  		 		 	   	  		 \end{array} \right)
\end{equation}
as a consequence of the underlying $S(3) \times S(3)$ symmetry. This would
imply that the basic flavor mixing matrix for leptons would not be close to
the unit matrix, as for the quarks, but close to the matrix
\begin{equation}
U = \left( \begin{array}{rrr}
\frac{1}{\sqrt{2}} & - \frac{1}{\sqrt{2}} & 0 \\
\frac{1}{\sqrt{6}} & \frac{1}{\sqrt{6}} & - \frac{2}{\sqrt{6}} \\
\frac{1}{\sqrt{3}} & \frac{1}{\sqrt{3}} & \frac{1}{\sqrt{3}}
\end{array} \right) \, ,
\end{equation}

i. e. there would be a qualitative difference between leptons and quarks.
Not taking into account contributions from the symmery breaking, the mixing
angles for solar neutrinos would be 45$^{\circ}$. However the symmetry breaking
might change this, and an angle like 35$^{\circ}$, in agreement with
observation, might result.

I am studying this at the moment. We would expect
$\Theta_{13} \approx - \frac{2}{\sqrt{6}} \sqrt{\frac{m_e}{m_{\mu}}}$,
which gives $sin^2 2 \Theta_{13} \approx 0.013$, too small to be seen now, but
possibly be seen in the forthcoming experiments.

King and McNamara looked at the group $SO(10) \times SU(3)$, with $SU(3)$
being some sort of flavor group for the three generations. The interplay between
flavor mixing, CP--violation and cosmology was studied by Marfatia. Okada talked
about the importance to look for the violation of lepton flavor in the
transition $\mu \rightarrow e \gamma, \mu A \rightarrow e^- A $ and $\tau
\rightarrow \mu \gamma $.

Ng discussed the importance of extra dimensions for
the neutrino physics. Here I have the problem to understand why neutrinos are
more sensitive towards new dimensions than the charged leptons or the quarks.

The r\^ole of leptogenesis, i. e. the generation of baryons via leptons, was
discussed by Kang. I still see many unknowns here and I am not convinced that
leptogenesis is the way to go.

The group $SO(10)$ was studied by Brachmacheri, in connection with
supersymmetry. Nevertheless, I should like to emphasize that
the unification based on $SO(10)$ has the advantage that one can achieve
unification also without supersymmetry, unlike the $SU(5)$--case.

In the talk of Kane we heared that much of flavor physics may only be
understood from string theory. I have problems with such statements. In string
theory there is hardly anybody really interested in flavor physics. Thus far
there is no progress in the understanding of flavor in string theory.

Berger discussed the possibility that the supersymmetric partner of the
$b$--quark is fairly light. I find this rather unlikely. Choi studied the
extra dimensions and interpreted them as the natural framework to understand
flavor and the fermion masses. I do not see a strong connection here.

The strong CP--problem was discussed by Chang. I like to emphasize that there
is no problem if e. g. the $u$--quark is massless. Thus the strong CP--problem
is connected with the thus far unresolved problem, how the quark masses are
generated. Only if we know more about this, we can see whether there is a
problem, or perhaps not.

In the talk of Oh, Baek and Park we heard again about the problem, related with
the decay $B \rightarrow \Phi K_s$ and $B \rightarrow \eta' K$. I think we
shall see that the Standard Model ist able to describe these decays, although
experimental errors might also play a r\^ole here.

Kang discussed the interesting relation $N_f = N_c$, in connection with the
``Little Higgs Model''. I do not wish to go in detail here, but the relation
$N_f = N_c$ is really quite remarkable. Nobody has a deep explanation of it,
but the physics in the future might give a simple reason for $N_f = N_c$.

Finally, let me stress that the physics of the future seems to be already
here, seen in the strange features of the quark--lepton mass spectrum and in
the flavor transtions.

For most theorists the way of the future is clear. Eventually we will discover
supersymmetry, and the road is open towards the superstring theory, perhaps
even towards new dimensions.

However I am not at all convinced that this is the way to go. At high energies
we might find new forms of substructure, perhaps also new types of forces.
The signals towards these new features might be just around the corner, like
new contributions to $g - 2$, or the decays $t \rightarrow c + $ glue,
$t \rightarrow c + \gamma$.

In QCD we understand the proton mass. It is proportional to $\Lambda_c$ and
related to the confinement aspect of QCD. In the Standard Model the $t$--mass
is given as the Yukawa coupling constant timed 246 GeV, the v.e.v. of the
$H$--field. But perhaps this mass reflects a new confinement phenomenon and
is related to a new kind of substrucutre. Only experiments in the TeV region
can tell us.

We need in theory new, predictive theories, which lead us beyond the
Standard Model. In flavor physics we are about at the same stage as particle
physics was in the beginning of the 60ies, before quarks were introduced,
before current algebra came, and before QCD came. Let us see what the future
has in store for us.

More definite answers will be given at the next meeting, the ICFP05. I wish
you a good return to your home institutes, and happy coming--back to I8CFP05.


\begin{thebibliography}{99}
\bibitem{CAL} X. Calmet and H. Fritzsch, Phys. Lett B496, 190 (2000)
\bibitem{Frit} H. Fritzsch and Z. Xing, Phys. Lett B372, 265--270 (1996)
\bibitem{Kob} M. Kobayashi and T. Maskawa, Prog. Theor. Phys. 49, 652 (1973)
\bibitem{Frit1} H. Fritzsch, Nucl. Phys. B155, 189 (1979)
\bibitem{Frit2} H. Fritzsch, Phys. Lett. B413, 396--404 (1997)
\end{thebibliography}
\end{document}